\documentclass[prb,twocolumn,preprintnumbers,superscriptaddress]{revtex4-1}

\usepackage{graphicx,epstopdf}
\usepackage{amssymb,amsmath}
\usepackage{dcolumn}
\usepackage{bm}
\usepackage{color}
\usepackage{enumerate}
\usepackage[colorlinks=true,citecolor=blue]{hyperref}
\hypersetup{colorlinks=true,citecolor=blue,linkcolor=blue,urlcolor=blue}

\usepackage{tabularx}
\usepackage{chemformula}

\usepackage{soul}

\begin{document}

\title{Theory of nonlinear excitonic response of hybrid organic perovskites in the regime of strong light-matter coupling}

\author{A. D. Belogur}
\affiliation{ITMO University, St. Petersburg 197101, Russia}

\author{D. A. Baghdasaryan}
\affiliation{Russian-Armenian University, Yerevan 0051, Armenia}

\author{I. V. Iorsh}
\affiliation{ITMO University, St. Petersburg 197101, Russia}

\author{I. A. Shelykh}
\affiliation{Science Institute, University of Iceland IS-107, Reykjavik, Iceland}
\affiliation{ITMO University, St. Petersburg 197101, Russia}

\author{V. Shahnazaryan}
\affiliation{ITMO University, St. Petersburg 197101, Russia}

\begin{abstract}
We present a quantitative study of the nonlinear optical response of layered perovskites placed inside planar photonic microcavities in the regime of strong light matter coupling, when excitonic and photonic modes hybridize and give rise to cavity polaritons. 
Two sources of nonlinearity are specified, the saturation of the excitonic transition with increase of the optical pump and Coulomb interaction between the excitons. 
It is demonstrated, that peculiar form of the interaction potential, specific to multilayer structure of organic perovskites, is responsible for substantial increase of the exciton binding energy and Rabi splitting  with respect to conventional semiconductor systems. This results in dominant contribution of the Rabi splitting quench effect in the nonlinear optical response.
Moreover, due to the tightly bound character of excitons, the density of Mott transition is essentially higher, allowing to reach extremely large polariton blueshifts of about 50 meV, which is order of magnitude higher than in conventional semiconductors.
\end{abstract}

\maketitle

\section{Introduction}

One of the most ambitious goals of the contemporary condensed matter physics is to study strongly-correlated matter. The onset of correlations is related to  the underlying nature of the quasiparticles, defining the material response. To observe  novel highly nonlinear physics, the unique combination of optical and electronic properties, and their fine interplay, is typically required. The latter becomes especially pronounced in the systems of reduced dimensionality and even more so in the hybrid low dimensional structures where quasiparticles of several types are present. Besides fundamental interest, the problem has strong application dimension, as strong nonlinear response is required for high-performance logical elements operating at quantum level. 

In this context, the systems with hybrid light-matter quasiparticles, such as planar microcavities in the regime of the strong light-matter coupling are of particular interest. The interaction between low dimensional excitons with cavity photons leads to the formation of cavity polaritons \cite{KavokinBook}, which possess the unique combination of extremely low effective mass and macroscopic coherent length, inherited from the photonic component of the polariton wavefunction, with strong polariton-polariton interactions stemming from their excitonic component. As a result, polariton systems form an attractive platform for the observation of plethora of quantum collective phenomena, which includes polariton lasing \cite{Kasprzak2006,Balili2007,Schneider2013,Ballarini2017}, polariton superfluidity,  \cite{Amo2009,Amo2011}, formation of solitons \cite{Sich2012,Hivet2012} and vortices \cite{Lagoudakis2008,Tosi2012,Cookson2021}, polarization multistability \cite{Gippius2007,Cerna2013,Gavrilov2013}, and nontrivial polariton lattice dynamics \cite{Ohadi2017,Gao2018,Pickup2020,Topfer2021}. These phenomena provide a solid basis for practical realization of ultra fast  polariton-based nonlinear optical integrated devices \cite{ShelykhReview,Amo2010,Liew2010,Ohadi2015,Dreismann2016,Opala2019,Jayaprakash2019}.

Sustained functioning of polariton based devices is defined by the excitonic binding energy $E_B$ and the Rabi splitting between the polariton modes $\Omega_R$, the latter being defined by the parameters of the cavity and optical oscillator strength of the excitonic transition. 
Since the first experimental observation of cavity polaritons in GaAs based samples, where both $E_B$ and $\Omega_R$ are about meV \cite{Weisbuch1992},  the search of the materials with more robust excitons and higher values of $\Omega_R$ continues. 
Among the perspective candidates, one should mention wide band semiconductors, such as GaN and ZnO \cite{Zamfirescu2002,Christopoulos2007,Li2013}, organic materials \cite{Lidzey1998,Kena2010,Kena2013,Plumof2014,Betzold2020,Yagafarov2020},  carbon nanotubes \cite{Graf2016,Mohl2018,Shahnazaryan2019} and monolayers of transition metal dichalcogenides (TMD) \cite{Schwarz2014,LiuMenon2015,Dufferwiel2015,Lundt2017,Schneider2018,Mortensen2020}.  

\begin{figure}
    \centering
    \includegraphics[width=0.99\linewidth]{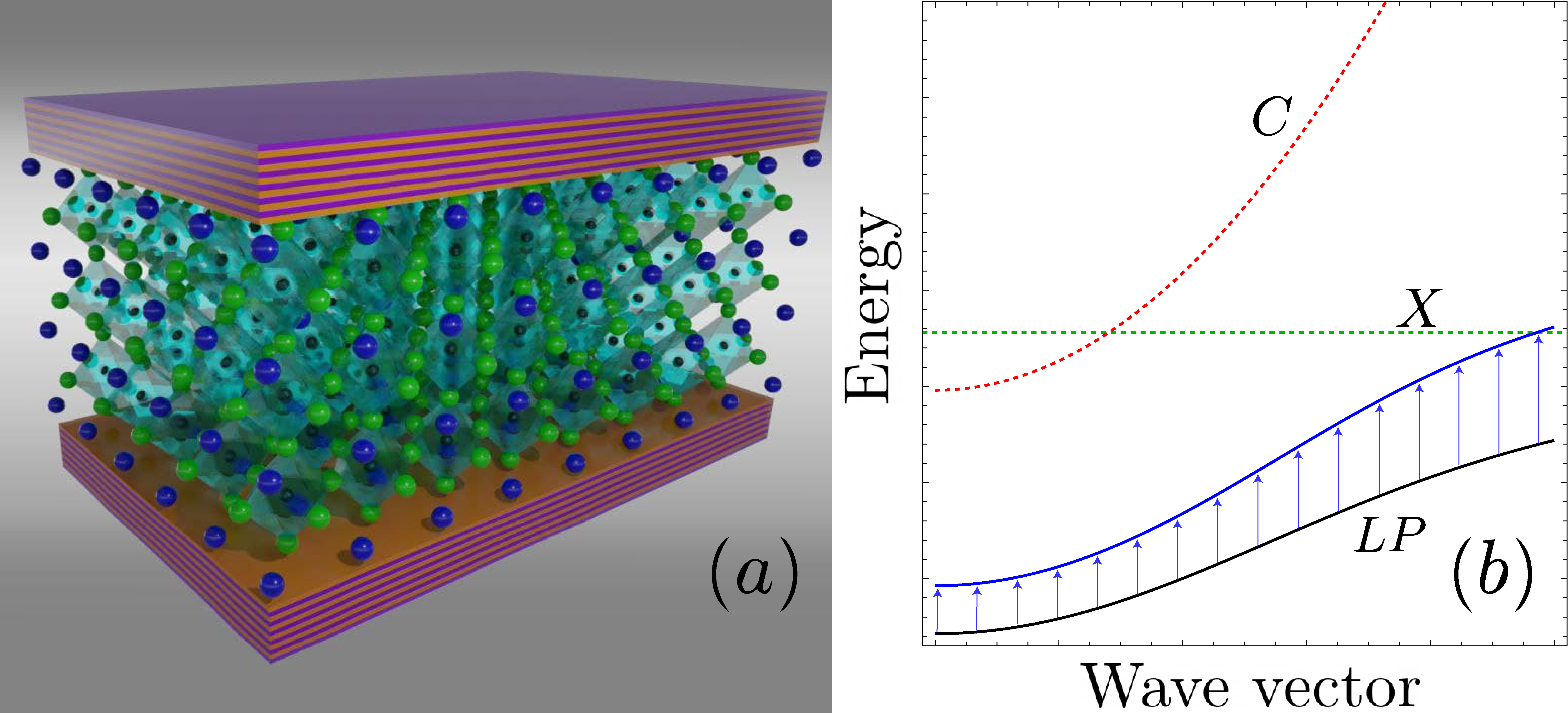}
    \caption{ 
    (a) The sketch of the system. 
    Multilayer 2D perovskite crystal is embedded in a microcavity represented by a pair of Bragg reflectors. 
    Due to the hybridization of the perovskite excitons with the cavity eigenmodes in the strong light-matter coupling regime  exciton-polaritons are formed.
    (b) The nonlinear spectrum of the polaritons. The photon (C) and exciton (X) modes in the strong coupling regime give rise to polaritons. 
    The energy of the lower polariton branch (LP) undergoes a blueshift  with increase of the excitation power due to the  many-body renormalization.
    }
    \label{fig:sketch}
\end{figure}

Recently, layered two-dimensional (2D) Ruddlesden-Popper  organic–inorganic metal halide perovskites (RPP) were presented as promising candidates for polaritonic applications \cite{Lanty2008,Su2017,Polimeno2020,Su2020}. 
The distinctive peculiarity of RPP is their multilayer structure, consisting of alternating layers of inorganic metal halides and organic media (see Fig.\ref{fig:sketch}). 
The number of halid layers, $n$ can vary from  $n=1$, which corresponds to atomically thin materials, to $n\rightarrow\infty$, characteristic to a 3D bulk material. 
The optical response of thin RPP films reveals robust excitonic peak up to room temperatures \cite{Ahmad2015,Straus2018,Li2019,Marongiu2019,Deng2020}. 
The corresponding exciton binding energy is up to 500 meV \cite{Tanaka2005,Yaffe2015,Mauck2019}, which is orders of magnitude higher, than in conventional semiconductor materials. 
The change of the number of layers strongly influences the excitonic properties, due to the modifications of the dielectric screening of electron-hole attraction, bandgap renormalization, and change in exciton-phonon coupling \cite{Saparov2016,Gong2018,Straus2018,Blancon2018,Quan2019}.

The tight binding between electrons and holes in perovskite materials leads to the giant values of the Rabi splitting in perovskite based microcavities (up to 270 meV \cite{Su2017}).
Robustness of perovskite polaritons allowed to achieve polariton BEC at room temperatures \cite{Su2020}. 
Moreover, the nonlinear optical response, measured as blueshift of exciton \cite{Huang2017,Abdelwahab2019,Ohara2019} or lower polariton mode \cite{Fieramosca2019}, was reported to be substantially higher, compared to conventional excitonic materials. 
This is surprising result. Indeed, it is generally believed, that in semiconductor microcavities the main contribution to the blueshift comes from exciton-exciton interactions \cite{Brichkin2011,Estrecho2019}, which is dominated by the electron and hole exchange  \cite{Ciuti1998,Tassone1999,Glazov2009}.
As overlap of the excitonic wavefunctions is required for this process, the interaction induced nonlinearity is stronger in materials with loosely bound excitons, such as InAs and GaAs, and should decrease with increase of the excitonic binding energy \cite{Shahnazaryan2017}. 

Later on, however, it was demonstrated, that this conclusion holds only for the case, when the interaction between the carriers is described by the Coulomb potential, and may change when the effects of the dielectric screening characteristic for 2D \cite{Keldysh1979} or layered \cite{Muljarov1995} systems, which lead to strong deviations of the interaction potential from simple $1/r$ scaling, are accounted for.  
Moreover, it was argued, that in the regime of strong light-matter coupling the effects of the saturation of the excitonic transition, leading to the quenching of the Rabi splitting, give important contribution to the blueshift \cite{Betzold2020,Yagafarov2020,Brichkin2011,Emmanuele2020}.
Therefore, detailed understanding of the microscopic origin of the giant nonlinear optical coefficients is still required for coherent description of complex optical phenomena observed in layered perovskites \cite{Ferrando2018}. 
Our work aims at bridging this evident gap. 
We consider a RPP material embedded in a planar cavity in the strong coupling regime (see Fig.~\ref{fig:sketch} (a)). 
We follow the approach developed in Refs.~\onlinecite{Muljarov1995,Suris2015} to define exciton binding energies and profiles of excitonic wavefunctions, which are then used to obtain the values of the Rabi splitting and nonlinearity coefficient, accounting for both exciton-exciton interaction and effects of the saturation of the excitonic transition.
The combination of both effects results in wavevector-dependent large optical nonlinearity of exciton-polaritons (Fig.~\ref{fig:sketch} (b)).


\section{Exciton states and exciton-exciton interaction}

\begin{figure}
    \centering
    \includegraphics[width=0.99\linewidth]{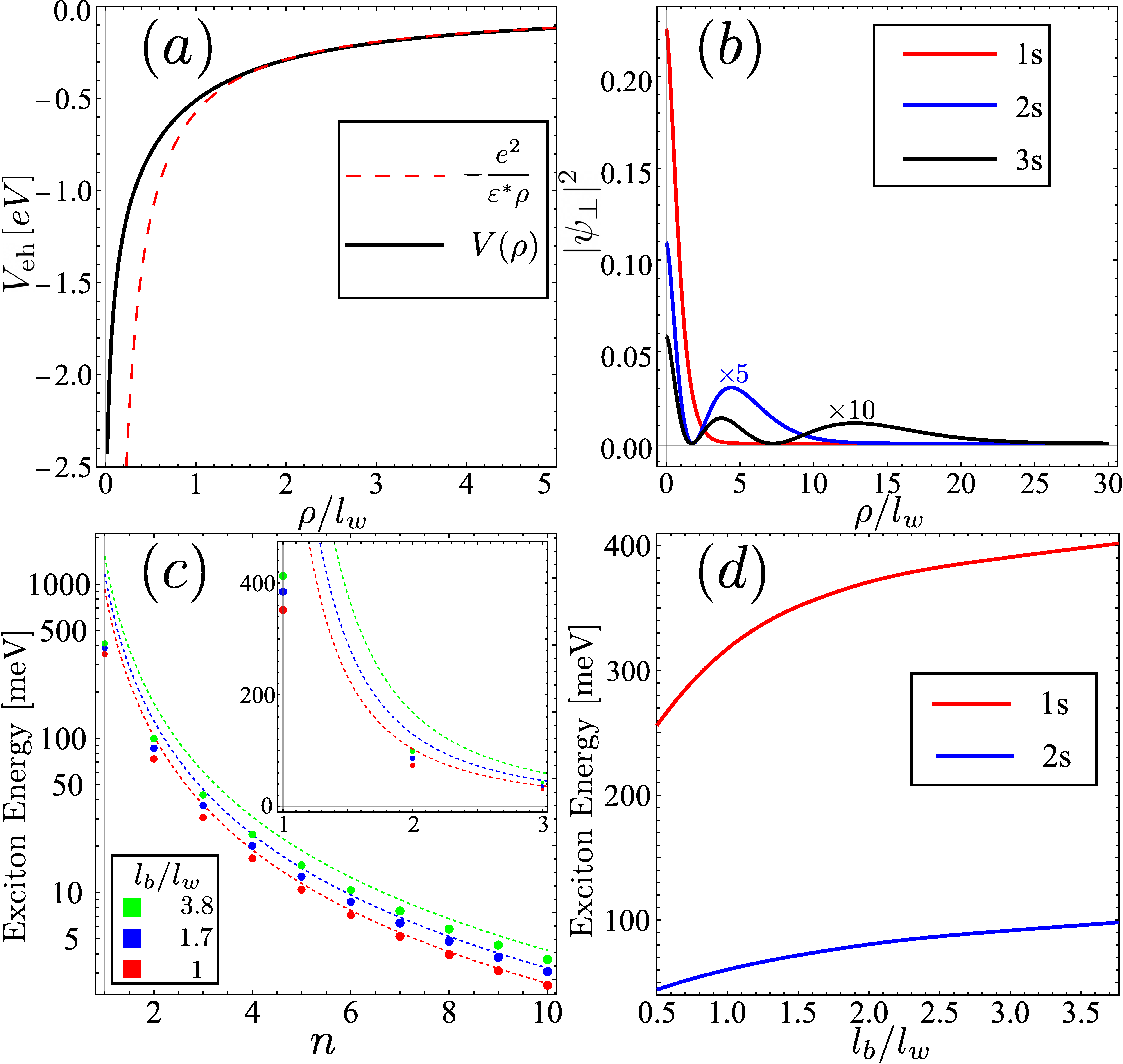}
    \caption{ 
    (a) Effective potential of the electron-hole in-plane Coulomb interaction. The red dashed curve corresponds to the conventional 2D Coulomb potential with $\propto 1/r$ scaling. 
    (b) Profiles of the in-plane wave functions of the exciton ground and excited states. 
    (c) Spectrum of the exciton energies for different values of an organic barrier length. The thin curves correspond to the  Rydberg-like spectra characteristic for the conventional Coulomb potential. The inset shows the strong deviation from Rydberg scaling for the lowest energy states.
    (d) The ground and excited state exciton energies versus the barrier to well width ratio.
    }
    \label{fig:exciton}
\end{figure}

We start with the determination of excitonic binding energy and the profile of the wavefunction of exciton states in considered system. 
As hybrid perovskites are direct bandgap materials, the ground exciton state is bright. 
The layered structure of material plays a role of a superlattice for charge carriers, with inorganic and organic parts corresponding to wall and barrier, respectively. 
As shown in Ref.~\cite{Muljarov1995}, this circumstance allows to separate the in-plane and normal-to-plane dynamics. The dynamics of charge carriers along the stacking direction is described by the following Schrodinger equation:
\begin{align}
    \label{eq:Schro_z}
    \left( - \frac{\hbar^2}{2m_{zi}} \frac{{\rm d}^2}{{\rm d} z_i^2} 
    + U_i (z_i)  \right) \psi ^i ( z_i)
    = E_z^i \psi^i( z_i),
\end{align}
where the index $i=e,h$ corresponds to electrons and holes, $m_{zi}$ are the particle effective masses along the stacking direction, and ${U_i(z_i)} = U^0_i(z_i) +  U^{\rm self}_i(z_i)$ 
is the single-particle confinement potential in conduction and valence bands. The term $U^0_i(z_i)$ corresponds to the bandgap mismatch in organic and inorganic layers, and $U^{\rm self}_i(z_i)$ is the self-energy correction due to the image charge effects. 
We assume that an electron and a hole are well localized and situated in the same layer.

The in-plane motion of an electron and a hole forming an exciton is predominantly defined by the shape of the attractive potential between them, which is strongly different from the Coulomb potential due to the layered structure of the material.
For excitonic wavefunction we use the ansatz, which allows for the separation of the relative and center of mass motions:
\begin{equation}
\Psi_{ {\bm Q}}(\bm{r_e},\bm{r_h}) = \psi^{e}(z_e) \psi^{h}(z_h) \psi_\perp(\rho;m) e^{i m\varphi} \frac{1}{\sqrt{A}} e^{i {\bm QR}} ,
\end{equation}
where $A$ is the normalization area, $\bm{R}=(m_e\bm{\rho_e}+m_h\bm{\rho_h})/(m_e+m_h)$ is the coordinate of exciton center of mass, with $m_e$, $m_h$ denoting the in plane components of the tensor of effective masses in the conduction and the valence bands.
$\bm{Q}$ is the wavevector of the center of mass, $\rho,\varphi$ are in-plane polar coordinates of the relative motion.
Restricting the consideration to $s-$ states only (m=0), we get:
\begin{align}
    \label{Schr_r}
    - \frac{\hbar ^2}{2\mu} \left( \frac{{\rm d}^2}{ {\rm d} \rho^2} + \frac{1}{\rho} \frac{{\rm d}}{{\rm d} \rho} \right) \psi_{\perp} + V_{eh}(\rho)  \psi_{\perp} = -E_b \psi_{\perp},
\end{align}
where $\mu=m_e m_h/(m_e+m_h)$ is the reduced mass, and $-E_b$ is the exciton binding energy. The effective electron-hole Coulomb interaction accounts for the averaging over $z-$ direction:
\begin{align}
    \label{V_eh}
    V_{eh} (\rho) =
    \int \left|\psi^e (z_e)\right|^2 \left|\psi^h(z_h)\right|^2
    U\left( z_e,z_h,\rho\right) {\rm d} z_e {\rm d} z_h,
\end{align}
where $U\left( z_e,z_h,\rho\right)$ is the bare interaction accounting for the image charge effects. The explicit expressions for $U\left( z_e,z_h,\rho\right)$, $U_e( z_e)$, $U_h( z_h)$ strongly depend on the geometry and the material composition, and are presented in the Appendix A.

For numerical calculations we consider the family of lead iodide RPPs, such as \ch{(C10H21NH3)2PbI4}.  The corresponding material parameters are shown in Table 1.
In Fig.~\ref{fig:exciton} (a) the radial dependence of the effective in-plane Coulomb interaction is presented. 
While at large distances it scales as $1/\rho$ typical for the Coulomb interaction, at smaller values $\rho$ it is essentially different, demonstrating a logarithmic scaling. 
This results in the deviation from the conventional Rydberg scaling of the exciton Bohr radius and the binding energy with the number of an excitonic state $n$.
Fig.~\ref{fig:exciton} (b) illustrates the spatial profile of the several lowest energy exciton states.
In Fig.~\ref{fig:exciton} (c) the energies of exciton states are presented. 
One clearly sees the spectrum is distinct from the Rydberg scaling at small values of principal quantum number $n$, as it is shown in the inset.
In Fig.~\ref{fig:exciton} (d) the energy of the exciton ground and first excited states versus the barrier length is shown. 
The enhancement of the exciton binding with increase of the barrier length is associated with the stronger localization of charge carriers within one layer.

We utilize the excitonic wavefunctions to evaluate the Kerr nonlinearity associated with the Coulomb interaction between excitons. 
It is characterized by several scattering  channels associated with direct and exchange processes \cite{Ciuti1998}. 
However, as in the case of conventional semiconductors the dominant contribution stems from the exchange interaction. 
The explicit expressions of the direct and exchange interaction matrix elements are presented in Appendix B. 

In Fig.~\ref{fig:XXint} (a) we present dependence of the exchange interaction of ground state excitons on the exchanged wavevector for several RPP materials. At small values of the wavevector $q$ the interaction is of repulsive character, and changes the sign at intermediate values $q$. 
This is in qualitative agreement with the previous studies of exciton-exciton interactions in conventional quantum wells \cite{Ciuti1998,Tassone1999,Glazov2009}. 
We fit the maximum of exchange interaction at zero wavevector as
\begin{equation}
V_{\rm exch} = \alpha E_b a_B^2,   
\end{equation}
where $a_B = \langle \psi_{\perp}| \rho | \psi_{\perp} \rangle$ is the exciton Bohr radius. The corresponding values of prefactor for different materials range in the region $\alpha \sim 3 \div 3.5 $, as shown in Table 2. Interestingly, this prefactor is different from both the cases of conventional semiconductor QWs, where $\alpha=6$ \cite{Tassone1999} and TMD monolayers where $\alpha\approx 2$ \cite{Shahnazaryan2017}. 
We further consider a hypothetical maximal blueshift due to exciton-exciton interaction, which can be defined as $\Delta E_{\rm m} = V_{\rm exch} n_{\rm m}$.  
Here $n_{\rm m} \approx 0.1 / a_B^2$ is the exciton density corresponding to Mott transition \cite{Mott1968}.
The corresponding values of blueshift, shown in Table 2, are in range $\Delta E_{\rm m} \sim 60 \div 110$ meV.
This is at least an order of magnitude larger compared to conventional 2D excitonic materials, such as GaAs \cite{Brichkin2011,Mukherjee2019,Estrecho2019} or GaN \cite{DiPaola2021}.
In Fig.~\ref{fig:XXint} (b) the direct interaction is presented. For all the considered materials it is essentially smaller compared with the exchange interaction, and has a characteristic dome-shape dependence on the exchanged wavevector \cite{Ciuti1998}.


\section{Strong light matter coupling in thin perovskite films}

We further consider the structure consisting of a thin film of RPP superlattice placed in the antinode of a planar cavity in the strong light-matter coupling regime, resulting in the formation of exciton polaritons.
The characteristic energy of the interaction of an exciton with cavity near-resonant eigenmode reads \cite{BursteinBook}:
\begin{equation}
    \label{eq:coupling}
    \Omega_0(k) =\sqrt{ \frac{E_C (k) N}{ \varepsilon \varepsilon_0 L_C} } |\psi_{\perp} (0)| d_{cv},
\end{equation}
where $N$ is the number of RPP layers, and $\varepsilon$ is the effective dielectric permittivity of a material of a cavity.
The photonic mode dispersion is $E_C(k) = E_C^0 + \hbar^2 k^2/(2m_C)$, with $E_C^0 = \pi \hbar c / (\sqrt{\varepsilon} L_C)$, $m_C=\pi \hbar / (c \sqrt{\varepsilon} L_C)$ denoting the cavity resonance and photonic effective mass. 
$L_C$ is the cavity length, $c$ is the speed of light, and $\psi_{\perp} (0)$ is the real space exciton wavefunction taken at $\bm{r}_e=\bm{r}_h$. 
Note, that due to the essential modulation of the Coulomb interaction here we cannot approximate the exciton wavefunction at the origin as  $\psi_{\perp}(0) = \sqrt{2/\pi} a_B^{-1}$.
$d_{cv}$ is the dipole matrix element for the optical interband transition, which is estimated applying a $k\cdot p $ perturbation method as
$d_{cv} \approx e \hbar /\sqrt{  \mu E_g }$.
The values of the exciton photon coupling $\Omega_0(0)$ are presented in Table 1.
As a reference we use $\Omega_0 (0) = 180$ meV, reported in Ref. \cite{Ouyang2020} for \ch{(BA)2(MA)3Pb4I13}, which we reporduce setting  $N = 14$ and using the corresponding material parameters.
We use the same value of $N$, treating thus  all the materials on an equal footing, and proceed with parameters of \ch{(C10H21NH3)2PbI4}, yielding in 
$\Omega_0 (0) = 285$ meV.

\begin{widetext}

\begin{table}[h!]\centering
\label{tab:data}
\begin{tabular}{ 
|c|c|c|c|c|c|c|c|c|c|c|c|c| }
\hline
\hline
Compound & $l_b$, & $l_w$, & $\varepsilon_b $ & $\varepsilon_w$ & $\mu, $ & $m^z_e,$ & $m^z_h,$  & $\Delta E_C$, & $\Delta E_V$, & $ E_b$,  & $a_B$, & $\Omega_0$  \\
$ $ & nm & nm & &  & $m_0$ & $m_0$ & $m_0$  &  eV & eV &  meV & nm  & meV  \\
\hline
\ch{(C10H21NH3)2PbI4} &  1.33  & 0.8   &2.1&6.05&0.17&0.2&0.5&1.91&1.91&352 &1.03&285 \\
\ch{(BA)2(MA)3Pb4I13} &   0.71  & 2.28   &2.23&5.19&0.196&0.35&0.55&2&3.1&206&1.28&180 \\
\ch{Y2PbI4}  &   0.82  & 0.8   &2.34&6.48&0.17&0.2&0.5&1.41&1.41&270&1.13&242 \\
\ch{Y2(CH3NH3)Pb2I7}    &   0.81  & 1.43   &2.34&6.48&0.17&0.2&0.5&1.41&1.41&188&1.39&185 \\
\hline
\end{tabular}
\caption{
Parameters of the representatives of the lead iodide family of RPP: organic spacer layer (barrier) and perovskite layer (well) thickness $l_{b,w}$ and dielectric constants $\varepsilon_{b,w}$, respectively, exciton in-plane reduced mass $\mu$ (units of $m_0$), electron and hole mass in $z$-direction $m^z_{e,h}$, respectively, conduction and valence band offsets $\Delta E_{C,V}$, respectively, exciton binding energy $ E_b$, and Bohr radius $ a_B$, light-matter coupling energy $\Omega_0$.
Here we used abbreviations \ch{Y}=\ch{C6H5C2H4NH3}, \ch{BA}=\ch{CH3CH2CH2CH2NH2}. }
\end{table}

\end{widetext}

\begin{table}[h!]\centering
\label{tab:data}
\begin{tabular}{ 
|c|c|c|c|c|c|c| }
\hline
\hline
Compound  &  $s$,& $ V$, & $\alpha$ & $\Delta E_{\rm m}$, & $s_2$, & $V^{(2)} $, \\
$ $  & nm$^2$  & $\mu$eV$\cdot \mu$m$^2$ & $ $ & meV & nm$^4$ & $\mu$eV$\cdot \mu$m$^4$ \\
\hline
\ch{(C10H21NH3)2PbI4} &  4.9&2.41 & 3.2 & 112.3 & -28.6&-1.2 $\cdot 10^{-5}$ \\
\ch{(BA)2(MA)3Pb4I13} &  7.4&2.06 & 3.1 & 63.7 & -65.9&-1.8 $\cdot 10^{-5}$ \\
\ch{Y2PbI4}  &5.8&2.41 & 3.5 & 94.2& -42&-1.3 $\cdot 10^{-5}$ \\
\ch{Y2(CH3NH3)Pb2I7}    &8.9&2.38 &  3.3 &62 & -95.7&-2.3 $\cdot 10^{-5}$ \\
\hline
\end{tabular}
\caption{
Parameters of the representatives of the lead iodide family of RPP associated with excitonic optical nonlinearity:  saturation rates of Rabi splitting $s$ and $s_2$, exciton-exciton Coulomb interaction rates $V$, $V^{(2)}$, the scaling of Coulomb interaction $\alpha$ and the maximal blueshift $\Delta E_{\rm m}$ (see the text).  }
\end{table}

\begin{figure}[h]
    \centering
    \includegraphics[width=0.99\linewidth]{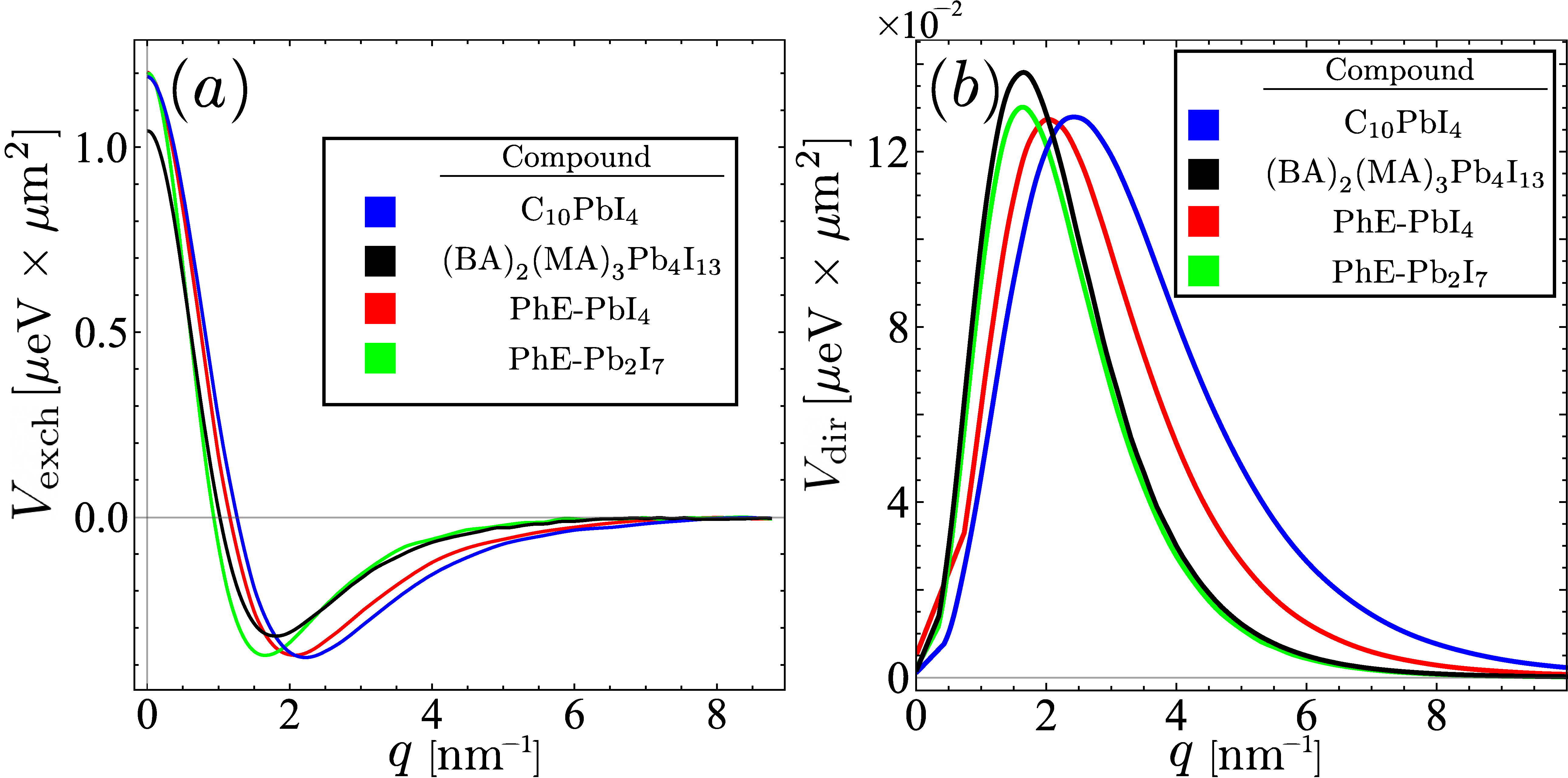}
    \caption{Exchange wavevector dependence of the exciton-exciton exchange (a) and direct( b)  interaction for several RPP materials.
    For all the materials the overall interaction is dominated by exchange scattering channel, and the contribution of the direct term becomes significant only at intermediate values of the wavevector.
    }
    \label{fig:XXint}
\end{figure}
\begin{figure}[h]
    \centering
    \includegraphics[width=0.99\linewidth]{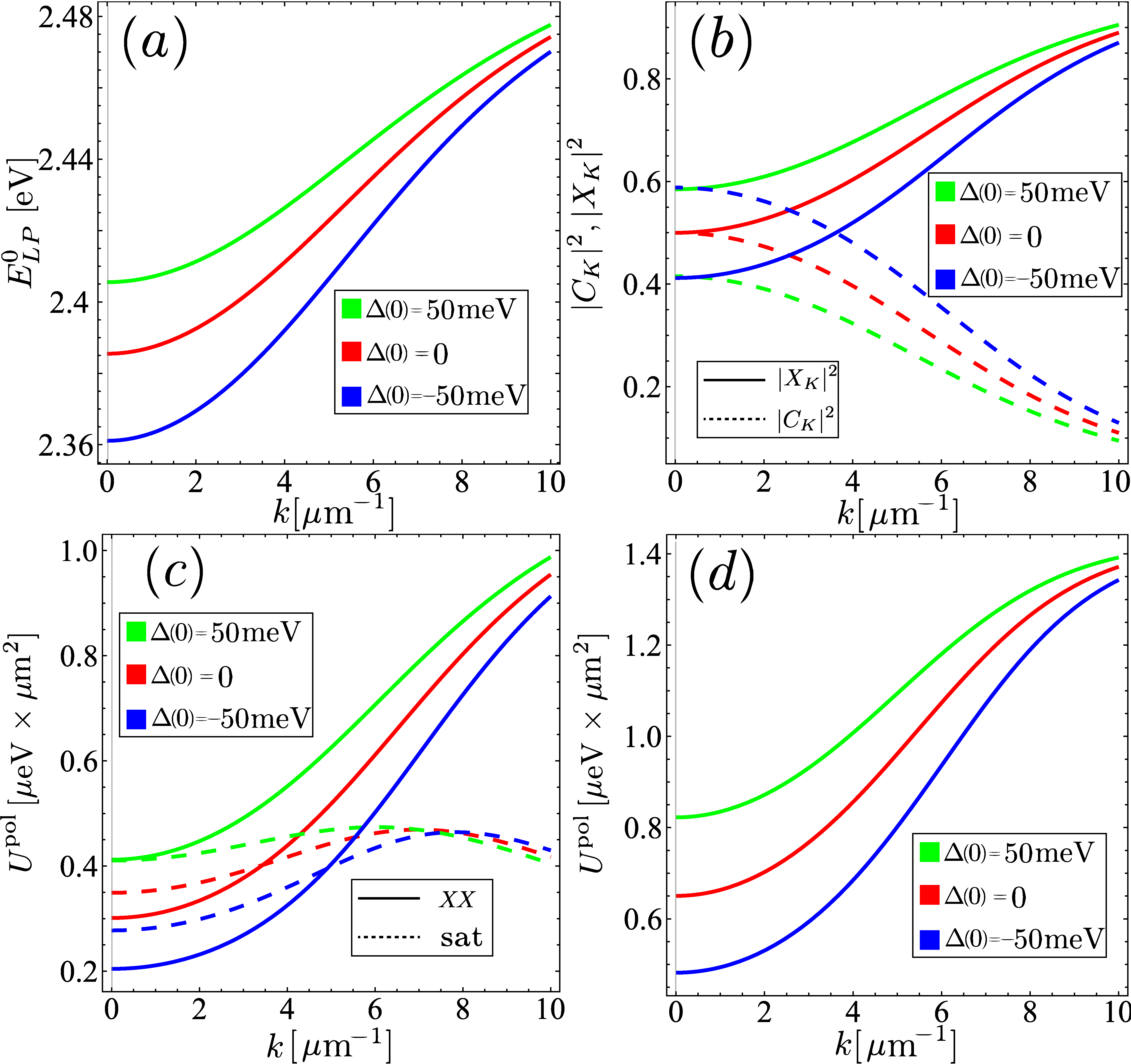}
    \caption{ (a) Dispersion of the lower polariton branch.  (b) The Hopfield coefficients at different detunings.
    (c) The wavevector dependence of Kerr nonlinearity associated with the Coulomb interactions (solid curves) and the quench of Rabi splitting (dashed curves). At small values of $k$ the quench of Rabi splitting gives dominant contribution due to the giant light-matter interaction. 
    At large values of the wavevector a lower polariton has exciton-like character, which defines the domination of the exciton-exciton interaction in this regime.
    (d) The momentum dependence of the total first order Kerr nonlinearity.
    }
    \label{fig:PLU1}
\end{figure}
\begin{figure}[h]
    \centering
    \includegraphics[width=0.99\linewidth]{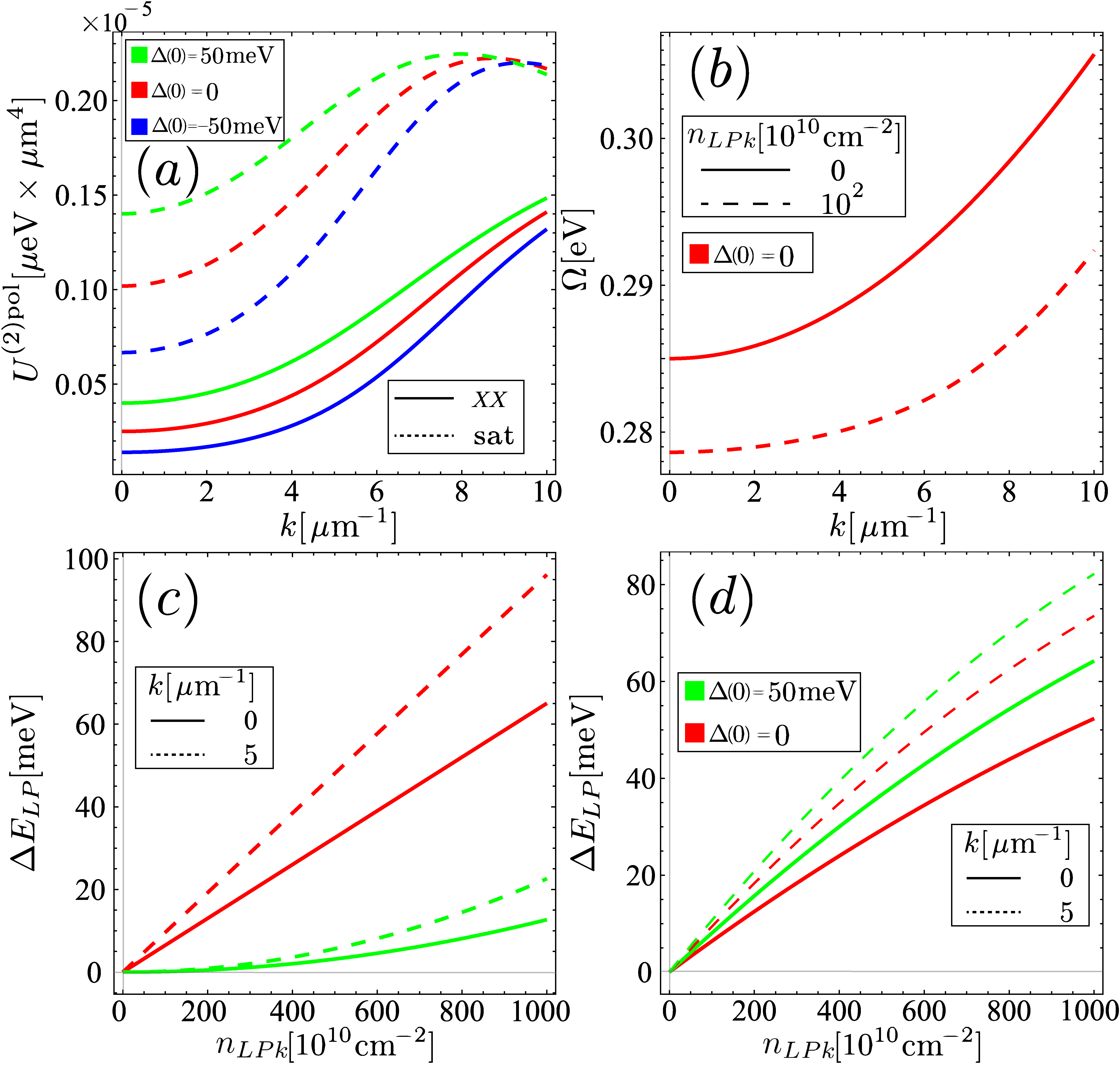}
    \caption{
    (a) The wavevector dependence of the second order Kerr nonlinearity associated with the Coulomb interactions (solid curves) and the quench of Rabi splitting (dashed curves).
    (b) The wavevector dependence of the light-matter coupling at small excitation (solid curve) and high particle densities (dashed curve).
    (c) The energy shifts stemming from the first (red curves) and the second order (green curves) nonlinearirity rates. Here $\Delta=0$. 
    For very large particle densities close to Mott transition density the second order contribution becomes comparable with the first order nonlinearity.
    (d) The density-dependent energy shift of the lower polariton branch for different values of the wavevector and different detunings. }
    \label{fig:PLnon}
\end{figure}

Using the couple oscillator model, the dispersion of the lower polariton branch can be found as \cite{KavokinBook}:
\begin{equation}
    \label{eq:Epolk}
    E_{LP}^0 (k) = \frac{1}{2} \left( E_C (k) + E_X - \sqrt{\Omega_0^2(k) + \Delta^2(k)} \right),
\end{equation} 
where $\Delta (k) = E_C(k) -E_X$ is the detuning between the exciton and the photon modes. Here $E_X = E_g - E_b$ is the exciton resonance energy, with $E_g$ denoting the single-particle bandgap.
As the effective mass of cavity photons is about 5 orders of magnitude smaller compared with  the effective mass of excitons, we can safely neglect the excitonic dispersion. Dispersion of the lower polariton branch for several values of the detuning is presented in Fig.~\ref{fig:PLU1} (a).

Due to  the small Bohr radius  of RPP excitons the density of Mott transition is 1-2 orders of magnitude larger compared to conventional semiconductor materials.
This circumstance allows to address the regime of elevated particle densities, where the strong interparticle correlations are well pronounced, and nonlinear terms providing the blueshift of polariton spectrum  are particularly important. 

One can identify two sources of the polariton blueshift. The first one is the Coulomb interactions between the excitons, already described above, and the second one is the reduction of the Rabi splitting due to the phase space filling effects, caused by the nontrivial quantum statistics of the excitons.
The contribution of both effects can be well quantified within the so-called coboson diagrammatic technique, developed by M. Combescot with co-authors \cite{CombescotReview}. 
Expanding the energy of lower polariton branch up to second order of polariton density $n_{LPk}$ one gets:
\begin{align}
    \label{eq:Epoltot}
    E_{LP}(k,n_{LPk}) =  E_{LP}^0 (k) + \Delta E_{LP}(k,n_{LPk}),
\end{align}
\begin{align}
    \label{eq:dEpoltot}
    \Delta E_{LP}(k,n_{LPk}) = U^{\rm pol} (k)  n_{LPk}-  U^{(2) \rm pol} (k)  n_{LPk}^2,
\end{align}
where the first term in Eq.~\eqref{eq:Epoltot} corresponds to polariton energy in low excitation regime, and nonlinear terms read as \cite{Carusotto2013}
(see also the Ref.~\cite{Emmanuele2020} for the derivation)
\begin{align}
    \label{eq:lpU1}
    U^{\rm pol} (k) =  U_{XX}^{\rm pol} (k) + U_{\rm sat}^{\rm pol} (k),
\end{align}
\begin{align}
    U^{(2) \rm pol} (k) =U_{XX}^{(2) \rm pol} (k) + U_{ \rm sat}^{(2) \rm pol} (k) , 
\end{align}
with 
\begin{align}
     U_{XX}^{\rm pol} (k) & \approx \frac{V}{2} |X_k|^4 , \\
     U_{\rm sat}^{\rm pol} (k) & = \Omega_0 (k) s X_k^3 C_k , 
\end{align}
\begin{align}
     U_{XX}^{(2) \rm pol} (k) & \approx  \frac{|V^{(2)}|}{6} |X_k|^6 ,  \\ 
     U_{\rm sat}^{(2) \rm pol} (k) & =  \Omega_0 (k) |s_2| X_k^5 C_k. 
\end{align}
Here $V \approx 2V_{\rm exch} (q=0)$ is the total exciton-exciton Coulomb interaction,
\begin{equation}
    C_k,X_k = \frac{1}{\sqrt{2}} \sqrt{ 1 \pm \frac{E_X - E_C(k)}{ \sqrt{(E_X - E_C(k))^2 + \Omega_0^2(k)} }  }
\end{equation}
are the Hopfield coefficients, and
\begin{align}
    s & = 2\frac{\sum\limits_{k} |\psi_k|^2 \psi_k }{\sum\limits_{q} \psi_q} , \\
    s_2 & = 2\frac{\sum\limits_{k} |\psi_k|^2 \psi_k \sum\limits_{k^\prime} |\psi_{k^\prime}|^4
    - \sum\limits_{k^{\prime \prime}} | \psi_{ k^{\prime \prime}} |^4 \psi_{k^{\prime \prime}} }
    {\sum\limits_{q} \psi_q}, 
\end{align}
are first and second order characteristic inverse saturation densities. $\psi_q$ denotes the exciton wavefunction in the inverse space. For hydrogen-like excitons these parameters read $s  = 16\pi a_B^2/7$ \cite{Brichkin2011}, and $s_2 = - 1152 \pi^2 a_B^4 / 455$ \cite{Emmanuele2020}.
The term $V^{(2)}$ corresponds to three-body Coulomb correlations  between excitons, with the explicit expression given in Appendix B.


In the Fig.~\ref{fig:PLU1} (c) the first order nonlinearity rates associated with  the exciton-exciton interaction and the saturation of Rabi splitting are shown.
Their relative importance depends on the Hopfield coefficients describing the percentage of the excitonic and photonic fractions, $X_k$ and $C_k$ respectively. 
These quantities are determined by the polariton wavevector $k$, and the detuning $\Delta$, as it is shown in Fig.~\ref{fig:PLU1} (b). 
At zero detuning and small $k$ the quench of the Rabi splitting dominates over the Coulomb scattering term. 
It is worth to note, that in the conventional semiconductor materials the Coulomb contribution is usually dominant \cite{Emmanuele2020,Shahnazaryan2020}, but for perovskites this is not the case because of the giant value of the Rabi splitting, $\Omega_0\sim$ 300 meV.
On the other hand, at larger values of the wavevector the photonic Hopfield coefficient rapidly decreases, and thus the Coulomb exchange starts to dominate. 
The Fig.~\ref{fig:PLU1} (d) illustrates the total first order nonlinearity, which enhances monotonously with the wavevector increase, given by the growth of exciton fraction in lower polariton branch.

In the Fig.~\ref{fig:PLnon} (a) the wavevector dependence of second order nonlinearities associated with  the exciton-exciton interaction and the saturation of Rabi splitting are presented. 
Remarkably, the latter is dominant in the wide range of the system parameters, which is again due to the giant light-matter coupling $\Omega_0$. 
In order to explicitly identify the impact of phase space filling on the Rabi splitting, we plot the  renormalized light-matter coupling, defined as $\Omega(k,n_{LPk}) = \Omega_0(k) - U_{\rm sat}^{\rm pol} (k) n_{LPk} + U_{\rm sat}^{(2) \rm pol} (k) n_{LPk}^2 $.
The corresponding wavevector dependence for small ($n_{LPk}=0$) and high excitation  ($n_{LPk}=10^{12}$ cm$^{-2}$) is depicted in Fig.~\ref{fig:PLnon} (b), demonstrating a significant quench of Rabi splitting.
In Fig.~\ref{fig:PLnon} (c) we present the energy shifts arising from the first and second order nonlinearity rates versus the polariton density. 
Evidently, the higher order correction becomes significant only at highly elevated densities $n_{LPk} \sim  10^{13}$ sm$^{-2}$.
Finally,  Fig.~\ref{fig:PLnon} (d) shows the total shift of lower polariton energy $\Delta E_{LP}$, defined by Eq.~\eqref{eq:dEpoltot}. 
One can immediately see for elevated particle densities the sublinear dependence of the blueshift on concentration, which is due to the negative sign of the second order correction. 
Another generic conclusion is that for polaritons containing larger fraction of excitons the blueshift is stronger.


\section{Conclusions} 
In conclusion, we presented the microscopic quantitative theory of excitonic nonlinear optical response of RPP materials in the regime of strong light-matter coupling, taking into account both exciton-exciton Coulomb exchange scattering and saturation of the exctionic optical transition. 
It was demonstrated, that due to the small Bohr radius of excitons the latter is dominant in a wide range of the system parameters
Moreover, the possibility to reach very large densities of exciton-polaritons allows to  drive the system in the highly nonlinear regime, characterized by giant blueshifts of polariton spectrum of the order of tens of meV, which is of order of magnitude higher than in conventional semiconductor microcavities.

This work was supported by the RA Science Committee and Russian Foundation for Basic Research (RF) in the frames of the joint research project SCS 20RF‐048 and RFBR 20-52-05005 accordingly.
I.A.S. and V.S. acknowledge support from the Icelandic Research Fund (project "Hybrid polaritonics"). V.S. thanks the University of Iceland for the hospitality.


\appendix

\section{Normal-to plane dynamics of charge carriers and the effective Coulomb interaction}

The charge carrier dynamics along the stacking direction are described by the single-particle potential ${U_i(z_i)} = U^0_i(z_i) +  U^{\rm self}_i(z_i)$. Here $U^0_i(z_i)$ corresponds to the difference of bandgap in organic (barrier) and inorganic (well) layers. 
We set $U^0_{e,h}(z_{e,h}) = 0 $ for the particle inside the well, and $U^0_{e}(z_{e}) = \Delta E_{C}$ and $U^0_{h}(z_{h}) = \Delta E_{V}$  for the particle inside the barrier.
The data for the values  $\Delta E_{C,V}$ for the different RPP compounds are presented in Table 1.
The self-energy correction $ U^{\rm self}_i(z_i)$, which is responsible for image charge effects, is defined by the expression \cite{Muljarov1995}:
\begin{align}
    U^{\rm self}_i(z_i) = \frac{e}{2}  \int_{0}^{\infty} \frac{q}{2\pi} \left[\phi(z_i,z_i,q) - \frac{2\pi e }{\varepsilon q} \right] {\rm d}q .
\end{align}
The form of the potential $\phi(z,z,q)$ and the value of the dielectric permittivity $\varepsilon$ depend on the particle location. In particular, inside the well
$\varepsilon = \varepsilon_w$ and 
\begin{align}
    \phi(z,z,q) = \frac{2 \pi e}{\varepsilon_b q} \frac{1}{\sinh[q_0]} \left\{\alpha^2 \sinh[q(l_b+l_w)]+ \right. \notag    \\
    \left. 
    \beta^2 \sinh[q(l_b-l_w)]+2\alpha\beta \sinh[q l_b ]\cosh[q(2z+l_w)] \right\},
\end{align}
where $\sinh[q_0] = ((\alpha^2 \cosh[q(l_b+l_w)]-$
$\beta^2 \cosh[q(l_b - l_w)])^2/\eta^2 - 1)^{1/2}$, $\eta = \varepsilon_b/ \varepsilon_w$,  $\alpha = (1 + \eta)/2$, $\beta = (1 - \eta)/2$. 
For the case particle is inside barrier one needs to substitute $ l_b \leftrightarrow l_w$, $\varepsilon_b \leftrightarrow \varepsilon_w$, $z \leftrightarrow -z$ and set $\varepsilon = \varepsilon_b$ .
The presented model leads to divergent behavior of potential at the interface of two adjacent layers. This problem is eliminated introducing a transition layer, where the potentials in two neighbouring layers are connected via a linear function. 

A bare two-particle interaction is defined by the expression:
\begin{align}
    U(z_e, z_h,\rho)  = - \frac{e}{2\pi}\int_{0}^{\infty} \phi(z_e,z_h,q) J_0(q\rho)q \, {\rm d} q .
\end{align}
%
For the case  particles are inside the well $-l_w < z_{e,h} < 0$ the interaction takes the form
\begin{align}
    \phi(z,z_0,q) = -\frac{2 \pi e}{\varepsilon_w q} \sinh(q |z-z_0|) + \frac{2 \pi e}{\varepsilon_b q \sinh[q_0]} \notag
    \\
    \left( 2\alpha\beta \sinh[q l_b] \cosh[q(z+z_0+l_w)]+(\alpha^2\sinh[q(l_b+l_w)]\right. \notag \\
    \left. +\beta^2 \sinh[q(l_b-l_w)])\cosh[q(z-z_0)] \right)
\end{align}
For the case where $0 < z_{e,h} < l_b$, one needs to substitute $l_b \leftrightarrow l_w$, $\varepsilon_b \leftrightarrow \varepsilon_w$, $z \leftrightarrow -z$, $z_0 \leftrightarrow -z_0$.

\section{The exciton-exciton interaction potential}

The theory of exciton-exciton Coulomb interaction between two-dimensional excitons was developed in Ref.~\cite{Ciuti1998}.
The corresponding rate of exciton-exciton interaction is defined as a matrix element of Coulomb interaction potential between charge carriers, corresponding to a process of elastic scattering between excitons with transfer of wave vector ${\bf q}$:
\begin{equation}
(\mathbf{Q}) + (\mathbf{Q}') \rightarrow (\mathbf{Q}+\mathbf{q})+(\mathbf{Q}'-\mathbf{q}),
\label{scattering}
\end{equation}
Here we expand the theory to account the different averaging of interaction components in $z-$ direction. 
In particular, the interaction potential reads as
\begin{align}
    V_I \left( {\bm \rho}_e, {\bm \rho} _{e^\prime}, {\bm \rho}_h, {\bm \rho}_{h^\prime} \right) = 
    &V_{ee}\left( \left| {\bm \rho}_{e^\prime} - {\bm \rho}_e \right| \right)
    + V_{hh}\left( \left| {\bm \rho}_{h^\prime} - {\bm \rho}_h \right| \right) \notag \\
    - &V_{eh}\left( \left| {\bm \rho}_{h^\prime} - {\bm \rho}_e \right| \right)
    - V_{he}\left( \left| {\bm \rho}_{e^\prime} - {\bm \rho}_h \right| \right),
\end{align}
where
\begin{align}
    \label{v_ee}
    V_{ij}\left( \left| {\bm \rho}_i - {\bm \rho}_j \right| \right) =    
     \int & V_{ij} \left(z_i, z_j, \left| {\bm \rho}_i - {\bm \rho}_j \right|  \right) \notag \\
    & \left| \psi_i^z( z_i) \right|^2 
    \left| \psi_j^z( z_j) \right|^2  {\rm d} z_i {\rm d} z_j .
\end{align}

The scattering amplitude of the process described by Eq.~(\ref{scattering}) is given by the matrix element
\begin{eqnarray}
V(\mathbf{Q},\mathbf{Q}',\mathbf{q})= V_{\mathrm{dir}}(\mathbf{Q},\mathbf{Q}',\mathbf{q}) +V^X_{\mathrm{exch}}(\mathbf{Q},\mathbf{Q}',\mathbf{q})+\notag\\ +V^e_{\mathrm{exch}}(\mathbf{Q},\mathbf{Q}',\mathbf{q})+V^h_{\mathrm{exch}}(\mathbf{Q},\mathbf{Q}',\mathbf{q})
\end{eqnarray}
where four terms correspond to direct interaction, exciton exchange, electron exchange, and hole exchange are given by the following expressions:
\begin{equation}
V_{\mathrm{dir}}(q) =  \frac{1}{A} [2 V_q \, g(\beta_e  q )  \,  g(\beta_h  q ) - V^{e}_q  g^2(\beta_h  q )  -  V^{h}_q g^2(\beta_e  q )  ],
\end{equation}
\begin{equation}
V^X_{\mathrm{exch}}(\Delta Q,q, \theta) =  V_{\mathrm{dir}}(\sqrt{(\Delta Q)^2 + q^2 - 2 q \Delta Q \cos \theta}),
\end{equation}
\begin{widetext}
\begin{equation}
V^e_{\mathrm{exch}}(\mathbf{Q},\mathbf{Q}',\mathbf{q})
= \int d^2\mathbf{r}_ed^2  \mathbf{r}_hd^2\mathbf{r}_{e'} d^2\mathbf{r}_{h'} 
\Psi^*_{\mathbf{Q}}(\mathbf{r}_e,\mathbf{r}_h) 
\Psi^*_{\mathbf{Q}'}(\mathbf{r}_{e'},\mathbf{r}_{h'})
V_{I}(\mathbf{r}_e,\mathbf{r}_h,\mathbf{r}_{e'},\mathbf{r}_{h'})
\Psi_{\mathbf{Q}+\mathbf{q}}(\mathbf{r}_{e'},\mathbf{r}_h)
\Psi_{\mathbf{Q}'-\mathbf{q}}(\mathbf{r}_e,\mathbf{r}_{h'}), 
\end{equation}
\begin{equation}
V^h_{\mathrm{exch}}(\mathbf{Q},\mathbf{Q}',\mathbf{q})
=\int d^2\mathbf{r}_ed^2 \mathbf{r}_hd^2\mathbf{r}_{e'}d^2\mathbf{r}_{h'}
\Psi^*_{\mathbf{Q}}(\mathbf{r}_e,\mathbf{r}_h)
\Psi^*_{\mathbf{Q}'}(\mathbf{r}_{e'},\mathbf{r}_{h'}) V_{I}(\mathbf{r}_e,\mathbf{r}_h,\mathbf{r}_{e'},\mathbf{r}_{h'}) 
\Psi_{\mathbf{Q}+\mathbf{q}}(\mathbf{r}_e,\mathbf{r}_{h'})
\Psi_{\mathbf{Q}'-\mathbf{q}}(\mathbf{r}_{e'},\mathbf{r}_h).
\end{equation}
\end{widetext}
Here $\Delta Q = |\mathbf{Q}' -\mathbf{Q}| $, and
\begin{align}
&g(\tau) =  2\pi \int J_0(\tau \rho)  |\psi_\perp^{\rm rad}(\rho)|^2 \rho {\rm d} \rho , \notag \\
&V_q = 2\pi \int J_0(q \tau)   V_{eh}(\tau)  \tau {\rm d} \tau , \notag\\
&V^{i}_q = 2\pi \int J_0(q \tau)   V_{ii}(\tau)  \tau {\rm d} \tau. 
\end{align}
We note that at $\Delta Q = 0$ one has $V^h_{\mathrm{exch}}(\mathbf{Q},\mathbf{Q}', 0) = V^e_{\mathrm{exch}}(\mathbf{Q},\mathbf{Q}', 0 )$.

The higher order Coulomb correlations are defined as \cite{Emmanuele2020}
\begin{align}
    \label{eq:Uexchq}
    & U^{(2)}  =   \sum_q |\psi_q|^4  \times \notag \\
    & \sum_{\vec{k},\vec{k}^\prime } 
    \psi^2_{ \bm{k}^\prime + \frac{  {\bm k} }{2} }
    \psi_{ \bm{k}^\prime - \frac{  {\bm k} }{2} }
    \left[
    2\psi_{ \bm{k}^\prime + \frac{   {\bm k} }{2} } V_{ {\bm k}}^{eh} 
    -\psi_{ \bm{k}^\prime - \frac{  {\bm k} }{2} }
    \left( V_{ {\bm k}}^{ee} + V_{ {\bm k}}^{hh} \right)
    \right] \notag \\
    - & \sum_{k,k^\prime}
    |\psi_k|^4    \left[  2 V_{{\bm k}-{\rm k}^\prime}^{eh} \psi_{k^\prime} \psi_k^* -|\psi_{k^\prime}|^2 
    \left(V_{{\bm k}-{\rm k}^\prime}^{ee} + V_{{\rm k}-{\bm k}^\prime}^{hh}\right)
     \right]
    .
\end{align}
%


\end{document}